\documentclass[
aps,%
12pt,%
final,%
notitlepage,%
oneside,%
onecolumn,%
nobibnotes,%
nofootinbib,% In the current version of REVTeX, when this option is on,
superscriptaddress,%
showpacs,%
showkeys,
centertags]%
{revtex4}
\usepackage{enumerate}
\usepackage{amssymb,relsize}
\usepackage{amsmath,amsfonts,amsthm,bm}
\usepackage{bbm}
\usepackage[force]{feynmp-auto}
    \usepackage[T1,T2A]{fontenc}
    \usepackage[english]{babel}

\def\be{\begin{equation}}
\def\ee{\end{equation}}

\def\sg{\sigma}

\def\bk{\hspace{-6.5pt}/}

\begin{document}

\title{Mass prediction for the last discovered member of the axial-vector nonet
with quantum numbers J$^{PC}=1^{+-}$}
\thanks{Corresponding author: mih@phys.uni-sofia.bg}

\author{M.V. Chizhov}
\affiliation{Department of Atomic Physics, Faculty of Physics, Sofia
University, 1164 Sofia, Bulgaria}
\author{M.N. Naydenov}
\affiliation{Department of Atomic Physics, Faculty of Physics, Sofia
University, 1164 Sofia, Bulgaria}

\begin{abstract}
In this paper we have shown that the novel mass relation among the
vector states, $\phi(1020)$ and $\phi(1680)$, with quantum numbers
J$^{PC}=1^{--}$ and axial-vector strangeonium state $h_1(s\bar{s})$
with quantum numbers $1^{+-}$ is also valid for nonzero current
quark mass. This relation predicts the mass of recently discovered
$h_1(s\bar{s})$ state by the Collaboration BESIII within
experimental accuracy.
\end{abstract}



\maketitle

\newpage
\section{Introduction}
It is known that there exist the two lowest-lying axial-vector
nonets of meson states with quantum numbers for their neutral
components J$^{PC}=1^{++}$ and $1^{+-}$. The first nonet, ``$A$'',
includes isotriplet of $a_1$ mesons, two isodoublets of $K_{1A}$
strange mesons and two isosinglet states $f_1$. The second nonet,
``$B$'', consists of isotriplet of $b_1$ mesons\footnote{The $b_1$
mesons consist of the light $u$ and $d$ quarks. Its name denotes the
type of the nonet and has no relation to $b$ quark.}, two
isodoublets of $K_{1B}$ strange mesons and two isosinglet states
$h_1$.

The two isosinglets of spin-1 states have nearly pure
$u\bar{u}+d\bar{d}$ and $s\bar{s}$ structures. There is still
discussion about identification of the two $f_1$ states among three
observed resonances $f_1(1285)$, $f_1(1420)$ and $f_1(1510)$. While
identification of the last from two $h_1(1170)$ and $h_1(1415)$
states has been established in last two years. The Collaboration
BESIII discovered axial-vector strangeonium state
$h_1(s\bar{s})$~\cite{BESIIIhep,BESIII} and confirmed its previous
observations by two collaborations LASS~\cite{LASS} and Crystal
Barrel~\cite{CB}.

There is interesting situation with the measured and the predicted
mass of the axial-vector strangeonium state with quantum numbers
$1^{+-}$. For the first time this state has been observed by the
Collaboration LASS in 1988 with mass $m^{\rm
LASS}_{h_1(s\bar{s})}=(1380\pm 20)$~MeV. Therefore, this state got
the name $h_1(1380)$. The latest measurements by the Collaboration
BESIII $m^{\rm BESIII}_{h_1(s\bar{s})}=(1423.2\pm 2.1\pm 7.3)$~MeV
and PDG average $m^{\rm PDG}_{h_1(s\bar{s})}=(1416\pm 8)$~MeV
\cite{PDG} required changing the name of this resonance to
$h_1(1415)$ in the last year. However, there was no accurate
prediction of the mass of this state among cited theoretical papers
in the first version of BESIII publication~\cite{BESIIIhep}. The
precise prediction of $h_1(s\bar{s})$ mass $m^{\rm
theor}_{h_1(s\bar{s})}=(1415\pm 13)$~MeV has been given in 2003 in
the paper~\cite{JETP}.

The detail consideration of this prediction and new investigation
will be presented below.

\section{The model}
Explanation of the spontaneous chiral symmetry breaking~\cite{Nambu}
and the introduction of quarks~\cite{GellMann,Zweig} give us a
principal possibility to describe the whole variety of the light
hadron states, in particular the quark-antiquark meson states. The
most theoretically and experimentally well studied is the nonet of
pseudoscalar mesons, which arise as pseudo-Goldstone bosons in
result of spontaneous and explicit chiral symmetry breaking. The
properties of pseudoscalar nonet can be accurately investigated
using Nambu -- Jona-Lasinio (NJL) model~\cite{NJL} or chiral
perturbation theory~\cite{ChPT}. Meanwhile, theoretical and
experimental situation with identification and explaning properties
of members of scalar nonet as quark-antiquark states is
unsatisfactory.

In this paper we consider a model for spin-1 nonets in the framework
of paper~\cite{JETP}. It is well known, that besides the two
considered lowest-lying axial-vector nonets, a nonet of vector
mesons exists. The latest consists of isotriplet $\rho$ mesons, two
isodoublets $K^*$ mesons and two isosinglets $\omega$ and $\phi$
with quantum numbers $1^{--}$, which have nearly pure
$u\bar{u}+d\bar{d}$ and $s\bar{s}$ structures, correspondingly. So,
there is an obvious {\em asymmetry} between the numbers of
axial-vector and vector nonets.

In paper \cite{JETP} in the framework of extended U(1) massless
quark NJL model a new approach to restoring the symmetry has been
suggested. In this way new mass relations among spin-1 mesons from
different nonets have been derived, which are confirmed
experimentally. The basic idea was in the consideration of all
possible Lorentz invariant {\em local} Yukawa interactions between
quark currents: $\overline{\psi}\psi$,
$\overline{\psi}\gamma^5\psi$, $\overline{\psi}\gamma_\mu\psi$,
$\overline{\psi}\gamma_\mu\gamma^5\psi$,
$\overline{\psi}\sigma_{\mu\nu}\psi$ and the corresponding meson
fields $S$, $P$, $V_\mu$, $A_\mu$, $T_{[\mu\nu]}$\footnote{The
notation $T_{[\mu\nu]}=\frac{1}{2}(T_{\mu\nu}-T_{\nu\mu})$ means
antisymmetrization on Lorentz indices.} with quantum numbers
$0^{++}$, $0^{-+}$, $1^{--}$, $1^{++}$, ($1^{--}$, $1^{+-}$). The
quark current $\overline{\psi}\sigma_{\mu\nu}\psi$ and the
corresponding second-rank antisymmetric tensor field $T_{[\mu\nu]}$
possess two types of different quantum numbers: $1^{--}$ and
$1^{+-}$. On the mass-shell $T_{[\mu\nu]}$ can be decomposed into
vector, $R_\mu$, and axial-vector, $B_\mu$, fields
\begin{equation}\label{decomposition}
    T_{[\mu\nu]}=(\hat{\partial}_\mu R_\nu-\hat{\partial}_\nu R_\mu)
    -\frac{1}{2}\,\epsilon_{\mu\nu\alpha\beta}
    (\hat{\partial}^\alpha B^\beta-\hat{\partial}^\beta B^\alpha)
\end{equation}
with corresponding quantum numbers $1^{--}$ and $1^{+-}$. Here we
introduce the definition
$\hat{\partial}_\mu=\partial_\mu/\sqrt{-\partial^2}$. Vice versa the
$R_\mu$ and $B_\mu$ fields can be expressed through $T_{[\mu\nu]}$:
\begin{equation}\label{RB}
    R_\mu=\hat{\partial}^\nu T_{[\mu\nu]},~~~~~
    B_\mu=\frac{1}{2}\,\epsilon_{\mu\nu\alpha\beta}\hat{\partial^\nu}T^{[\alpha\beta]},
\end{equation}
which due to antisymmetry of $T_{[\mu\nu]}$ obey the obvious
identities
\begin{equation}\label{RBconservation}
    \partial^\mu R_\mu\equiv 0,~~~~~~
    \partial^\mu B_\mu\equiv 0.
\end{equation}

Therefore, the existing $h_1(s\bar{s})$ meson with quantum numbers
$1^{+-}$ from axial-vector ``$B$'' nonet can be described by
axial-vector field $B_\mu$ and requires the inclusion in the model
the quark current $\overline{\psi}\sigma_{\mu\nu}\psi$ and the
corresponding second-rank antisymmetric tensor field $T_{[\mu\nu]}$.
Since the two vector fields $V_\mu$ and $R_\mu$ have the same
quantum numbers $1^{--}$ they could mix, which leads to the two
physical states, $\phi$ and $\phi'$ for U(1) $s\bar{s}$ quark
structure. So, extending this suggestion to U(3) model, it can be
proposed that the lowest-lying vector nonet of physical states with
quantum numbers $1^{--}$ is produced from a mixing of vector and
tensor nonets, similar to the two $K_1(1270)$ and $K_1(1400)$
physical states, which are superpositions of $K_{1A}$ and $K_{1B}$
states from the corresponding axial-vector nonets.

The new relation between the masses of $\phi(1020)$,
$\phi'=\phi(1680)$ and $h_1(s\bar{s})$ physical states~\cite{JETP}
\begin{equation}\label{massrel}
    2m^2_\phi-m_\phi m_{\phi'}+2m^2_{\phi'}=3m^2_{h_1(s\bar{s})},
\end{equation}
which predicts the mass of undiscovered then $h_1(s\bar{s})$ state,
has been obtained in approximation of zero current quark mass.
However, it can be rather applied to $\omega(782)$,
$\omega'=\omega(1650)$ and $h_1(1170)$ physical states, which
consist of the light $u$ and $d$ quarks. Therefore, to confirm the
relation (\ref{massrel}), here we investigate the case of nonzero
current quark mass $m_0$.

In order to obtain relation (\ref{massrel}) we consider only
interactions of vector, $V_\mu$, tensor, $T_{[\mu\nu]}$, and scalar,
$S$, fields with the field of strange quark, $\psi$. Linearized NJL
Lagrangian with auxiliary (without kinetic terms) boson fields, has
the form
\begin{eqnarray}
{\cal L}_0=\overline{\psi}(q\bk -m_0) \psi
&+&g_S\,\overline{\psi}\psi\;S-\frac{\mu^2_S}{2}S^2
+g_V\,\overline{\psi}\gamma^\mu\psi\;V_\mu +\frac{\mu^2_V}{2}V_\mu^2
\nonumber
\\
&+&\frac{g_T}{2}\,\overline{\psi}\sg^{\mu\nu}\psi\;T_{[\mu\nu]}
+\frac{\mu^2_T}{4}\left(4\,\hat{\partial}^\mu T_{[\lambda\mu]}
\hat{\partial}_\nu T^{[\lambda\nu]}-T_{[\mu\nu]}T^{[\mu\nu]}\right),
\label{yukawa}
\end{eqnarray}
where $\mu^2_S$, $\mu^2_V$ and $\mu^2_T$ are bare mass
terms\footnote{The form of the mass term for the antisymmetric
tensor field is suggested in \cite{MPLA}.}. Using relations
(\ref{decomposition},\ref{RB},\ref{RBconservation}) the last line in
eq.~(\ref{yukawa}) can be rewritten as
\begin{equation}\label{T2RB}
g_T\,\overline{\psi}\sg^{\mu\nu}\psi\;\hat{\partial}_\mu R_\nu
+ig_T\,\overline{\psi}\sg^{\mu\nu}\gamma^5\psi\;\hat{\partial}_\mu
B_\nu + \frac{\mu_T^2}{2}\left(R^2_\mu+B^2_\mu\right).
\end{equation}
The corresponding Feynman diagrams are presented in Fig.~\ref{fig1}.

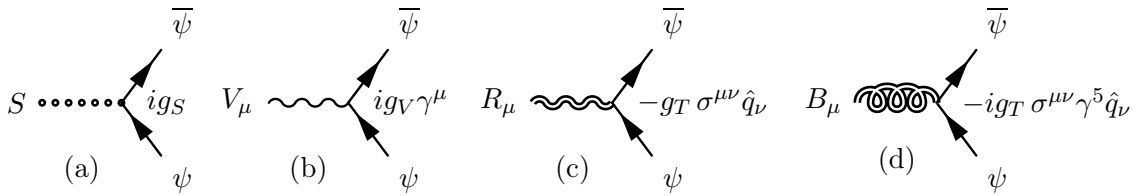
\vspace{0.5cm}
\begin{figure}[h]
      \centering

\begin{fmffile}{diagrams1}

\begin{fmfgraph*}(50,40)
\fmfleft{i1} \fmfright{o1,o2}
\fmf{dbl_dots,label=(a),l.d=0.4w}{i1,v1} \fmf{fermion}{o1,v1,o2}
 \fmflabel{$\psi$}{o1}
 \fmflabel{$\overline{\psi}$}{o2}
 \fmflabel{$S$}{i1}
 \fmflabel{$~ig_S$}{v1}
\end{fmfgraph*}
\hspace{10mm}
\begin{fmfgraph*}(50,40)
\fmfleft{i1} \fmfright{o1,o2} \fmf{boson,label=(b),l.d=0.4w}{i1,v1}
\fmf{fermion}{o1,v1,o2}
 \fmflabel{$\psi$}{o1}
 \fmflabel{$\overline{\psi}$}{o2}
 \fmflabel{$V_\mu$}{i1}
 \fmflabel{$~ig_V\gamma^\mu$}{v1}
\end{fmfgraph*}
\hspace{15mm}
\begin{fmfgraph*}(50,40)
\fmfleft{i1} \fmfright{o1,o2}
\fmf{dbl_wiggly,label=(c),l.d=0.4w}{i1,v1} \fmf{fermion}{o1,v1,o2}
 \fmflabel{$\psi$}{o1}
 \fmflabel{$\overline{\psi}$}{o2}
 \fmflabel{$R_\mu$}{i1}
 \fmflabel{$~-g_T\,\sigma^{\mu\nu}\hat{q}_\nu$}{v1}
\end{fmfgraph*}
\hspace{23mm}
\begin{fmfgraph*}(50,40)
\fmfleft{i1} \fmfright{o1,o2}
\fmf{dbl_curly,label=(d),l.d=-0.55w}{v1,i1} \fmf{fermion}{o1,v1,o2}
 \fmflabel{$\psi$}{o1}
 \fmflabel{$\overline{\psi}$}{o2}
 \fmflabel{$B_\mu$}{i1}
 \fmflabel{$~-ig_T\,\sigma^{\mu\nu}\gamma^5 \hat{q}_\nu$}{v1}
\end{fmfgraph*}
 \end{fmffile}
 \caption{\label{fig1}Feynman diagrams for Yukawa interactions. Here $\hat{q}_\mu=q_\mu/\sqrt{q^2}$,
where $q_\mu$ is incoming four-momentum of $R_\mu$ and $B_\mu$
states.}
\end{figure}

\section{Quantum corrections and symmetry breaking}
\begin{figure}[h]
      \centering

\begin{fmffile}{diagrams2}

\begin{fmfgraph*}(60,40)
\fmfleft{i}\fmf{dbl_dots,tension=1}{i,v1}
\fmfright{o}\fmf{dbl_dots,tension=1}{v2,o}
\fmf{fermion,left,tension=0.2,label=(a),l.d=0.1w}{v2,v1}
\fmf{fermion,left,tension=0.2}{v1,v2}
\end{fmfgraph*}
\hspace{15mm}
\begin{fmfgraph*}(70,60)
\fmfleft{i1}\fmf{dbl_dots,tension=2}{i1,v1}
\fmfright{o1,o2}\fmf{dbl_dots,tension=2}{v2,o1}
\fmf{dbl_dots,tension=2}{v3,o2}
\fmf{fermion,label=(b),l.d=.4w}{v2,v1} \fmf{fermion}{v1,v3,v2}
\end{fmfgraph*}
\hspace{15mm}
\begin{fmfgraph*}(80,60)
\fmfleft{i1,i2}\fmf{dbl_dots,tension=2}{i1,v1}\fmf{dbl_dots,tension=2}{i2,v2}
\fmfright{o1,o2}\fmf{dbl_dots,tension=2}{v4,o1}\fmf{dbl_dots,tension=2}{v3,o2}
\fmf{fermion,label=(c),l.d=-.3w}{v4,v1} \fmf{fermion}{v1,v2,v3,v4}
\end{fmfgraph*}
\end{fmffile}
 \caption{\label{fig2}Quantum corrections to the self-energy part (a) and
the self-interactions (b,c) of the scalar fields.}
\end{figure}
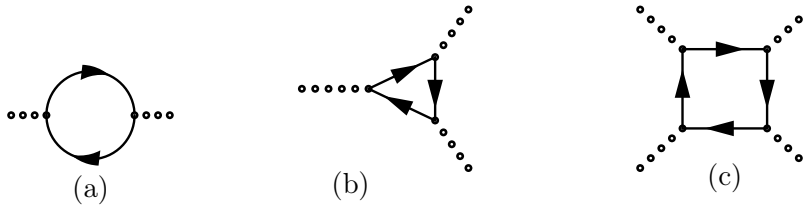

Kinetic terms for the meson fields can be obtained through radiative
quantum corrections. Let's calculate one-loop self-energy
contributions for all the boson fields. For example, the
contribution from the self-energy diagram of the scalar field
(Fig.~\ref{fig2}a) has the form

\vspace{-1cm}
\begin{eqnarray}\label{PS}
 \Pi^{SS}(q)&=&ig_S^2 N_C \int \frac{{\rm d}^4p}{(2\pi)^4}
 {\rm Tr}\left[\left(p\bk-m_0\right)^{-1}\left(p\bk - q\bk-m_0\right)^{-1}\right]
 \nonumber\\
 &=&4ig_S^2 N_C\int \frac{{\rm d}^4p}{(2\pi)^4}\frac{1}{p^2-m_0^2}
 -2ig_S^2 N_C\int \frac{{\rm d}^4p}{(2\pi)^4}\frac{q^2-4m^2_0}{\left(p^2-m_0^2\right)^2}
 +{\rm finite~terms}\nonumber\\
 &=&4g_S^2 N_C I_2-8g_S^2 N_C m_0^2 I_0+2g_S^2 N_C I_0~q^2
 +{\rm finite~terms},
\end{eqnarray}
where
\begin{equation}\label{I2}
  I_2\equiv i\int \frac{{\rm d}^4p}{(2\pi)^4}\frac{1}{p^2-m_0^2}=
  \int \frac{{\rm d}^4p_E}{(2\pi)^4}\frac{1}{p_E^2+m_0^2}>0
\end{equation}
is the quadratically divergent integral and
\begin{equation}\label{I0}
  I_0\equiv -i\int \frac{{\rm d}^4p}{(2\pi)^4}\frac{1}{\left(p^2-m_0^2\right)^2}=
  \int \frac{{\rm d}^4p_E}{(2\pi)^4}\frac{1}{\left(p_E^2+m_0^2\right)^2}>0
\end{equation}
is the logarithmically divergent integral.

The first two terms in the last line of eq.~(\ref{PS}) give
contribution into the mass term of the scalar field
$m^2_S=\mu_S^2+4g_S^2 N_C(2m_0^2 I_0-I_2)$. The third term indicates
the appearance of the kinetic term. To normalize the scalar field
wave function correctly, we must require fulfillment of the
condition
\begin{equation}\label{norm}
  2g_S^2N_C I_0=1.
\end{equation}

Applying the similar procedure to all boson fields introduced in
(\ref{yukawa}) and (\ref{T2RB})
\begin{equation}\label{PVV}
    \Pi^{VV}_{\mu\nu}=\frac{4}{3}\,g_V^2 N_C I_0 (q_\mu q_\nu-q^2 g_{\mu\nu})
    -2g_V^2 N_C (m_0^2 I_0+I_2)g_{\mu\nu}
    +{\rm finite~terms},
\end{equation}
\begin{equation}\label{PVR}
    \Pi^{VR}_{\mu\nu}=4g_V g_T N_C I_0 \frac{m_0}{\sqrt{q^2}}
    (q_\mu q_\nu-q^2 g_{\mu\nu})+{\rm
    finite~terms},\mbox{\hspace{3cm}}
\end{equation}
\begin{equation}\label{PRR}
    \Pi^{RR}_{\mu\nu}=\frac{2}{3}\,g_T^2 N_C I_0 (q_\mu q_\nu-q^2 g_{\mu\nu})
    +4g^2_T N_C I_0 \frac{m^2_0}{q^2}
    (q_\mu q_\nu-q^2 g_{\mu\nu})+{\rm finite~terms},
\end{equation}
\begin{equation}\label{PBB}
    \Pi^{BB}_{\mu\nu}=\frac{2}{3}\,g_T^2 N_C I_0 (q_\mu q_\nu-q^2 g_{\mu\nu})
    -4g^2_T N_C I_0 \frac{m^2_0}{q^2}
    (q_\mu q_\nu-q^2 g_{\mu\nu})+{\rm finite~terms},
\end{equation}
the following useful relations among various Yukawa coupling
constants from normalization of kinetic terms are obtained:
\begin{equation}\label{relations}
  3g_S^2=2g_V^2=g_T^2=\frac{3}{2N_C I_0}.
\end{equation}
In other words, due to the dynamic origin of the kinetic terms, all
interactions in the model are described by only one coupling
constant, for example, $g=g_S$.

\begin{figure}[h]
      \centering

\begin{fmffile}{diagram3}

\begin{fmfgraph*}(70,60)
\fmfleft{i1}\fmf{dbl_plain,tension=2}{i1,v1}
\fmfright{o1,o2}\fmf{dbl_plain,tension=2}{v2,o1}
\fmf{dbl_dots,tension=2}{v3,o2} \fmf{fermion}{v2,v1,v3,v2}
\fmfv{label=$\langle S\rangle$,decor.shape=cross,decor.angle=45}{o2}
\end{fmfgraph*}
\hspace{15mm}
\begin{fmfgraph*}(80,60)
\fmfleft{i1,i2}\fmf{dbl_plain,tension=2}{i1,v1}\fmf{dbl_dots,tension=2}{i2,v2}
\fmfright{o1,o2}\fmf{dbl_plain,tension=2}{v4,o1}\fmf{dbl_dots,tension=2}{v3,o2}
\fmf{fermion}{v1,v2,v3,v4,v1} \fmfv{label=$\langle
S\rangle$,decor.shape=cross,decor.angle=45}{o2,i2}
\end{fmfgraph*}
\hspace{20mm}
\begin{fmfgraph*}(70,60)
\fmfleft{i1}\fmf{dbl_wiggly,tension=2}{i1,v1}
\fmfright{o1,o2}\fmf{dbl_curly,tension=2}{v2,o1}
\fmf{dbl_dots,tension=2}{v3,o2} \fmf{fermion}{v2,v1,v3,v2}
\fmfv{label=$\langle S\rangle$,decor.shape=cross,decor.angle=45}{o2}
\end{fmfgraph*}

\end{fmffile}
 \caption{\label{fig3} Quantum contributions to the mass terms and mixing
between vector bosons after spontaneous symmetry breaking. Here the
double lines correspond simultaneously to any of the $R_\mu$ or
$B_\mu$ bosons.}
\end{figure}
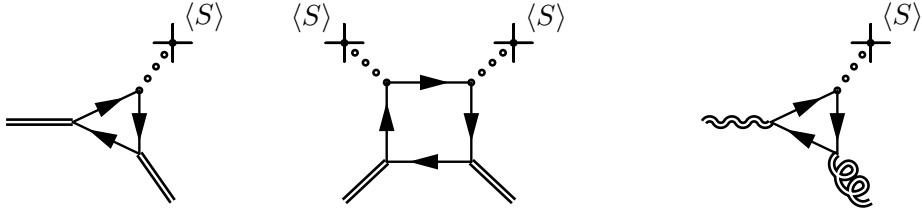

The self-interactions in Fig.~\ref{fig2}(b,c) lead to the following
effective potential for the scalar field
\begin{equation}\label{Veff}
    V_{\rm eff}=\frac{m^2_S}{2}\,S^2-2g m_0 S^3+\frac{g^2}{2}S^4.
\end{equation}
The extremum condition
\begin{equation}\label{dV}
    \left.\frac{{\rm d}V_{\rm eff}}{{\rm d}S}\right|_{S=\langle
    S\rangle}
    =m^2_S\langle S\rangle-6gm_0\langle S\rangle^2+2g^2\langle S\rangle^3=0
\end{equation}
at negative $ m^2_S$ leads always to the absolute minimum of the
effective potential with nontrivial solution $g\langle
S\rangle=(3m_0-\sqrt{9m^2_0-2 m^2_S})/2<0$. It corresponds to the
spontaneous symmetry breaking, which provides the strange quark with
positive constituent mass
\begin{equation}\label{ms}
    m_s=m_0-g\langle S\rangle>0.
\end{equation}
It also leads to additional contributions to mass terms and mixing
between vector bosons (Fig.~\ref{fig3}).

There are no additional contribution to the mass of the vector field
$V_\mu$, $m^2_V=\mu^2_V-2g_V^2 N_C (m_0^2 I_0+I_2)$, while its
mixing (\ref{PVR}) with the vector field $R_\mu$ gets such
contribution after symmetry breaking
\begin{equation}\label{dPVR}
    \Delta\Pi^{VR}_{\mu\nu}=-4g_V g_T N_C I_0 \frac{g\langle S\rangle}{\sqrt{q^2}}
    (q_\mu q_\nu-q^2 g_{\mu\nu})+{\rm finite~terms}
\end{equation}
in such a way that it depends only on physical constituent mass of
the strange quark:
\begin{equation}\label{PVRfull}
    \Pi^{VR}_{\mu\nu}+\Delta\Pi^{VR}_{\mu\nu}= \sqrt{\frac{18
    m^2_s}{q^2}}\,
    (q_\mu q_\nu-q^2 g_{\mu\nu})+{\rm finite~terms}.
\end{equation}
In its turn $R_\mu$ and $B_\mu$ fields get additional contributions
\begin{eqnarray}
\label{dPRR}
  \Delta\Pi^{RR}_{\mu\nu} &=& 4g^2_T N_C I_0
  \frac{g^2\langle S\rangle^2-2m_0g\langle S\rangle}{q^2}
    (q_\mu q_\nu-q^2 g_{\mu\nu})+{\rm finite~terms}, \\\label{dPBB}
  \Delta\Pi^{BB}_{\mu\nu} &=& -4g^2_T N_C I_0
  \frac{g^2\langle S\rangle^2-2m_0g\langle S\rangle}{q^2}
    (q_\mu q_\nu-q^2 g_{\mu\nu})+{\rm finite~terms},
\end{eqnarray}
to their mass terms $m^2_R=\mu^2_T-4g^2_T N_C I_0 (m^2_0+g^2\langle
S\rangle^2-2m_0 g\langle S\rangle)=\mu^2_T-6m^2_s$ and
$m^2_B=\mu^2_T+4g^2_T N_C I_0 (m^2_0+g^2\langle S\rangle^2-2m_0
g\langle S\rangle)=\mu^2_T+6m^2_s$, which also depend on the
physical constituent mass of strange quark.

As a result the effective Lagrangian for the spin-1 bosons has the
form
\begin{equation}\label{L1}
    {\cal L}_{\rm eff}=-\frac{1}{2}\left(V_\mu~R_\mu\right)
    \begin{pmatrix}
    q^2-m^2_V & \sqrt{18 m^2_s q^2}\\
    \sqrt{18 m^2_s q^2} & q^2-m^2_B+12m^2_s\\
    \end{pmatrix}\!
    \begin{pmatrix}
    V^\mu\\ R^\mu\\
    \end{pmatrix}
    -\frac{1}{2}\,B_\mu\!\left(q^2-m^2_B\right)\!B^\mu.
\end{equation}
In paper \cite{JETP} it has been shown that the masses of spin-1
isovector states $\rho$, $\rho'=\rho(1450)$, $b_1$ and spin-1 light
isosinglets $\omega$, $\omega'$, $h_1(1170)$ correspond to the
hypothesis of the maximal mixing between $V_\mu$ and $R_\mu$ states.
It leads to the additional constraint
\begin{equation}\label{maxVR}
    m^2_V=m^2_B-12m^2_s.
\end{equation}
Zeros of determinant of the matrix between $(V_\mu~R_\mu)^T$
doublets correspond to the masses for the physical $\phi$ and
$\phi'$ bosons, while $h_1(s\bar{s})$ meson has $m_B$ mass. Using
Vieta's formulas
\begin{equation}\label{Vieta}
    m^2_\phi+m^2_{\phi'}=2(m^2_V+9m^2_s),~~~~~
    m_\phi m_{\phi'}=m^2_V
\end{equation}
for quadratic equation $(q^2)^2-2(m^2_V+9m^2_s)q^2+m^4_V=0$ and
relation (\ref{maxVR}) we have reproduced the mass formula
(\ref{massrel}) without the assumption of initial massless quarks.

\section{Conclusion}
In this paper we have shown that novel mass relation (\ref{massrel})
for U(1) case of the strange quark with nonzero current mass, $m_0$,
has the same form as for initially massless quark. Therefore, the
prediction of the mass of axial-vector strangeonium state
$h_1(s\bar{s})$ with quantum numbers $1^{+-}$ in \cite{JETP} is
valid. This is also confirmed experimentally by the Collaboration
BESIII~\cite{BESIII}.

%\pagebreak

\end{document}